\documentstyle[epsfig]{elsart}
\newcommand{\lsim}{\mathrel{\rlap{\lower4pt\hbox{\hskip0pt$\sim$}}
\raise1pt\hbox{$<$}}}
\newcommand{\sfrac}[2]{\mbox{\footnotesize $\frac{#1}{#2}$}}
\begin{document}
\begin{frontmatter}
 \hspace*{\fill}{Preprint Numbers: \parbox[t]{100mm}{ANL-PHY-8753-TH-97
        \hspace*{\fill} nucl-th/9707003\\
        KSUCNR-103-97}}

\title{Pion mass and decay constant}
\author[anl]{Pieter Maris,}
\author[anl]{Craig D. Roberts}
\author[ksu]{and Peter C. Tandy}
\address[anl]{Physics Division, Bldg. 203, Argonne National Laboratory,\\
Argonne IL 60439-4843, USA}
\address[ksu]{Centre for Nuclear Research, Department of Physics, Kent State
University, Kent OH 44242}
\begin{abstract}
Independent of assumptions about the form of the quark-antiquark scattering
kernel we derive the explicit relation between the pion Bethe-Salpeter
amplitude, $\Gamma_\pi$, and the quark propagator in the chiral limit;
$\Gamma_\pi$ necessarily involves a non-negligible $\gamma_5\gamma\cdot P$
term ($P$ is the pion four-momentum).  We also obtain exact expressions for
the pion decay constant, $f_\pi$, and mass, both of which depend on
$\Gamma_\pi$; and demonstrate the equivalence between $f_\pi$ and the pion
Bethe-Salpeter normalisation constant in the chiral limit.  We stress the
importance of preserving the axial-vector Ward-Takahashi identity in any
study of the pion itself, and in any study whose goal is a unified
understanding of the properties of the pion and other hadronic bound states.
\end{abstract}
\begin{keyword}
Goldstone Bosons; Dynamical Chiral Symmetry Breaking; Dyson-Schwinger
equations; Nonperturbative QCD\\[2mm] 
\end{keyword}
\end{frontmatter}
In the strong interaction spectrum the pion is identified as both a Goldstone
mode, associated with dynamical chiral symmetry breaking (D$\chi$SB), and a bound
state composed of $u$- and $d$-quarks.  This dichotomy is interesting
because, while $m_\rho/2 \simeq m_N/3\equiv M_q$: the constituent-quark mass,
$m_\pi/2$ is only $\approx 0.2 M_q$; i.e., the pion is very much less-massive
than other comparable strong interaction bound states.  The constituent-quark
mass, $M_q\simeq 350\,$MeV, provides an estimate of the ``effective mass'' of
quarks bound in a hadron; and the ratio of this to the
renormalisation-group-invariant current-quark mass of $u$- and $d$- quarks
($\hat m \sim 10\,$MeV) indicates the magnitude of nonperturbative
dressing effects on light-quark propagation characteristics.  The particular
nature of the pion can be represented by the question: ``How does one form an
almost-massless bound state from very massive constituents {\it without}
fine-tuning?''  The answer to this question is at the core of an
understanding of D$\chi$SB in terms of the elementary degrees of freedom in QCD.

In addressing this question we employ the Dyson-Schwinger
equations~\cite{dserev} (DSEs), which provide a nonperturbative,
renormalisable, continuum framework for analysing quantum field theories, and
adapt the discussion of Ref.~\cite{JJ73}.  We begin with the renormalised,
homogeneous, pseudoscalar Bethe-Salpeter Equation (BSE)\footnote{We employ a
Euclidean space formulation with
$\{\gamma_\mu,\gamma_\nu\}=2\delta_{\mu\nu}$, $\gamma_\mu^\dagger =
\gamma_\mu$ and $a\cdot b=\sum_{i=1}^4 a_i b_i$.  A spacelike vector,
$k_\mu$, has $k^2>0$.}
\begin{eqnarray}
\label{genbse}
\left[\Gamma_\pi^j(k;P)\right]_{tu} &= & 
\int^\Lambda_q  \,
[\chi_\pi^j(q;P)]_{sr} \,K^{rs}_{tu}(q,k;P)\,,
\end{eqnarray}
where $k$ is the relative and $P$ the total momentum of the quark-antiquark
pair, $\chi_\pi^j(q;P) \equiv S(q_+) \Gamma_\pi^j(q;P) S(q_-)$,
$r$,\ldots,$u$ represent colour, flavour and Dirac indices, $q_\pm=q\pm P/2$,
and $\int^\Lambda_q \equiv \int^\Lambda d^4 q/(2\pi)^4$ represents
mnemonically a {\em translationally-invariant} regularisation of the
integral, with $\Lambda$ the regularisation mass-scale.  The final stage of
any calculation is to remove the regularisation by taking the limit $\Lambda
\to \infty$.  In (\ref{genbse}), $S$ is the dressed-quark propagator and $K$
is the fully-amputated quark-antiquark scattering kernel; the important
features of both are discussed below.

The homogeneous BSE is an eigenvalue problem.  Solutions exist only for
particular, separated values of $P^2$; and the eigenvector associated with
each eigenvalue, the Bethe-Salpeter amplitude (BSA): $\Gamma(k;P)$, is the
one-particle-irreducible, fully-amputated quark-meson vertex.  In the
isovector, pseudoscalar channel the solution associated with the lowest
eigenvalue is the pion.  This solution of (\ref{genbse}) has the general
form~\cite{LS69}
\begin{eqnarray}
\label{genpibsa}
\Gamma_\pi^j(k;P) & = &  \tau^j \gamma_5 \left[ i E_\pi(k;P) + 
\gamma\cdot P F_\pi(k;P) \rule{0mm}{5mm}\right. \\
\nonumber
& & \left. \rule{0mm}{5mm}+ \gamma\cdot k \,k \cdot P\, G_\pi(k;P) 
+ \sigma_{\mu\nu}\,k_\mu P_\nu \,H_\pi(k;P) 
\right]\,.
\end{eqnarray}

In Ref.~\cite{DS79} a ``Feynman-like'' gauge is employed: $D_{\mu\nu}(k)
\propto \delta_{\mu\nu}$, in which case the Dirac algebra entails $H_\pi
\equiv 0$.  The general defects of such an Ansatz are discussed in
Ref.~\cite{sep}.  Herein we note only that the Slavnov-Taylor identity:
$k_\mu D_{\mu\nu}=k_\mu D_{\mu\nu}^{\rm free}(k)$, states that the
longitudinal part of the dressed-gluon propagator is {\it independent} of
interactions and hence, even in Feynman gauge, $D_{\mu\nu}(k) \not\propto
\delta_{\mu\nu}$ and $H_\pi \not\equiv 0$.  The interplay between $F_\pi$,
$G_\pi$, $H_\pi$ and $F_R$, $G_R$, $H_R$, discussed below in connection with
(\ref{bwti})-(\ref{gwti}), is also overlooked in Ref.~\cite{DS79}.

Important in (\ref{genbse}) is the renormalised dressed-quark propagator,
which is obtained from the quark DSE
\begin{eqnarray}
\label{gendse}
S(p)^{-1} & = & Z_2 (i\gamma\cdot p + m_{\rm bm})
+\, Z_1 \int^\Lambda_q \,
g^2 D_{\mu\nu}(p-q) \gamma_\mu S(q)
\Gamma_\nu(q,p) \,,
\end{eqnarray}
where $D_{\mu\nu}(k)$ is the renormalised dressed-gluon propagator,
$\Gamma_\mu(q;p)$ is the renormalised dressed-quark-gluon vertex and $m_{\rm
bm}$ is the current-quark bare mass that appears in the Lagrangian.  The
solution of (\ref{gendse}) has the general form
\begin{equation}
\label{sinvp}
S(p)^{-1} = i \gamma\cdot p A(p^2) + B(p^2)\,.
\end{equation}

In (\ref{gendse}), $Z_1$ and $Z_2$ are, respectively, the renormalisation
constants for the quark-gluon vertex and quark wave function.\footnote{In
discussing renormalisation we follow the conventions of
Ref.~\protect\cite{pt84}} Each is fixed by the requirement that the
associated Schwinger function (vertex, propagator) take a prescribed value at
the renormalisation point, $p^2=\mu^2$, large and spacelike.  For example,
$Z_2$ is fixed by requiring that $A(\mu^2)$ take a prescribed value.  As with
each renormalisation constant herein, they depend on the renormalisation
point and the regularisation mass-scale.  For example, at one-loop in QCD,
$ Z_2 = [
\ln(\mu^2/\Lambda_{\rm QCD}^2)
/\ln(\Lambda^2/\Lambda_{\rm QCD}^2)]^{\gamma_S}$,
where $\mu$ is the renormalisation point and the anomalous dimension is
$\gamma_S = - 2 \xi /(33 - 2 N_f)$, with $\xi$ the gauge parameter and $N_f$
the number of quark flavours.  Also the renormalised current-quark mass is
$m_{(\mu)} \equiv Z_2 Z_4^{-1} m_{\rm bm}\equiv Z_m^{-1} m_{\rm bm}$, which
yields $m_{(\mu)} = \hat m /[\sfrac{1}{2} \ln(\mu^2/\Lambda_{\rm
QCD}^2)]^{\gamma_m}$, where $\hat m$ is the renormalis\-ation-group-invariant
current-quark mass and $\gamma_m = 12/(33 - 2 N_f)$.  We
note that: at one-loop $Z_2\equiv 1$ in Landau gauge, $\xi=0$; the anomalous
dimension, $\gamma_m$, of the mass renormalisation constant,
$Z_m = [
\ln(\mu^2/\Lambda_{\rm QCD}^2)
/\ln(\Lambda^2/\Lambda_{\rm QCD}^2)]^{\gamma_m}$, 
is independent of $\xi$ to all orders in perturbation theory; and the chiral
limit is defined by\footnote{The arguments presented herein cannot be applied
in a straightforward fashion to models whose ultraviolet behaviour is that of
quenched QED$_4$, such as Ref.~\cite{FR96}, where the chiral limit can't be
defined in this way.  The difficulties encountered in such cases are
illustrated in Ref.~\cite{hsw97}.}  $\hat m = 0$.

The quark condensate: $\langle \bar q q\rangle \propto \int^\Lambda_q {\rm
tr}[S(q)]$, is an order parameter for D$\chi$SB and (\ref{gendse}) has been
used extensively to study this phenomenon.  As summarised in
Ref.~\cite{dserev}, using any form of the gluon propagator that is strongly
enhanced in the neighbourhood of $k^2=0$; i.e., in the infrared, consistent
with the results of Refs.~\cite{mrp,gpropir}, and with any vertex that is
free of kinematic singularities~\cite{mrp,vertex}, the quark DSE admits a
nonzero solution for $B(p^2)$ in the chiral limit; i.e., one has D$\chi$SB
{\it without} fine-tuning.  In addition, any model study that is able to
provide a quantitatively good description of observables; such as: hadron
masses, decay constants, scattering lengths, etc., involves the generation of
a Euclidean constituent-quark mass: $M^E_{u,d} \approx 400\,$MeV; i.e., the
generation of a large ``effective mass'' for the $u$- and
$d$-quarks\footnote{$M^E_f$ is defined~\cite{sep} as the solution of: $p^2
A_f(p^2)^2 = B_f(p^2)^2$, where $f$ labels the quark flavour.  This is not
the quark pole mass, which need not exist in a theory with confinement.  The
ratio $M_f^E/\hat m_f$ is a single, indicative and quantitative measure of
the nonperturbative effects of gluon-dressing on the quark propagator.}.
This is illustrated, for example, in Refs.~\cite{FR96,dcsb}.  A large quark
``effective mass'' is therefore a direct and unavoidable consequence of the
infrared enhancement of $D_{\mu\nu}(k)$.

As a final introductory remark, we note that the renormalised quark-antiquark
scattering kernel in (\ref{genbse}), $K^{rs}_{tu}(q,k;P)$, which also appears
implicitly in (\ref{gendse}) because it is the kernel in the inhomogeneous
integral equation satisfied by $\Gamma_\mu(q;p)$, is the sum of a countable
infinity of skeleton diagrams.\footnote{By definition, $K$ does not contain
quark-antiquark to single gauge-boson annihilation diagrams, such as would
describe the leptonic decay of the pion, nor diagrams that become
disconnected by cutting one quark and one antiquark line.}  This is why we
speak of model studies above: any quantitative study of (\ref{genbse}) and
(\ref{gendse}) necessarily involves a truncation of this kernel whose
reliability cannot be gauged {\it a priori}$\,$.  It is, however, incumbent
upon practitioners to avoid drawing conclusions that are artefacts of a given
truncation.  No truncation of the kernel is employed in deriving the general
results presented herein.

In considering the pion, understanding chiral symmetry, and its explicit and
dynamical breaking, is crucial.  These features are expressed in the
axial-vector Ward-Takahashi identity (AV-WTI), which involves the isovector
axial-vector vertex:
\begin{eqnarray}
\label{genave}
\left[\Gamma_{5\mu}^j(k;P)\right]_{tu} & = &
Z_A \, \left[\gamma_5\gamma_\mu \frac{\tau^j}{2}\right]_{tu} \,+
\int^\Lambda_q \, [\chi_{5\mu}^j(q;P)]_{sr} \,K^{rs}_{tu}(q,k;P)\,,
\end{eqnarray}
$\chi_{5\mu}^j(q;P) \equiv S(q_+) \Gamma_{5\mu}^j(q;P) S(q_-)$, that has the
general form
\begin{eqnarray}
\label{genavv}
\Gamma_{5 \mu}^j(k;P) & = &
\frac{\tau^j}{2} \gamma_5 
\left[ \gamma_\mu F_R(k;P) + \gamma\cdot k k_\mu G_R(k;P) 
- \sigma_{\mu\nu} \,k_\nu\, H_R(k;P) 
\right]\\
&+ & \nonumber
 \tilde\Gamma_{5\mu}^{j}(k;P) 
+ \frac{P_\mu}{P^2 + m_\phi^2} \phi^j(k;P)\,,
\end{eqnarray}
where $F_R$, $G_R$, $H_R$ and $\tilde\Gamma_{5\mu}^{i}$ are regular as
$P^2\to -m_\phi^2$, $P_\mu \tilde\Gamma_{5\mu}^{i}(k;P) \sim {\rm O }(P^2)$
and $\phi^j(k;P)$ has the structure depicted in (\ref{genpibsa}).  By
convention, the renormalisation constant, $Z_A$, is chosen so as to fix the
value of $F_R(k;P=0)|_{k^2=\mu^2}$.  This form admits the possibility of at
least one pole term in the axial-vector vertex but does not require it.

Substituting (\ref{genavv}) into (\ref{genave}) and equating putative pole
terms, it is clear that, if present, $\phi^j(k;P)$ satisfies (\ref{genbse}).
Since (\ref{genbse}) is an eigenvalue problem that only admits a
$\Gamma_\pi^j \neq 0$ solution for $P^2= -m_\pi^2$, it follows that
$\phi^j(k;P)$ is nonzero only for $P^2= -m_\pi^2$ and the pole mass is
$m_\phi^2 = m_\pi^2$.  Hence, if $K$ supports such a bound state, the
axial-vector vertex contains a pion-pole contribution whose residue, $r_A$,
is not fixed by these arguments; i.e.,
\begin{eqnarray}
\label{truavv}
\Gamma_{5 \mu}^j(k;P) & = &
\frac{\tau^j}{2} \gamma_5 
\left[ \gamma_\mu F_R(k;P) + \gamma\cdot k k_\mu G_R(k;P) 
- \sigma_{\mu\nu} \,k_\nu\, H_R(k;P) \right]\\
&+ & \nonumber
 \tilde\Gamma_{5\mu}^{i}(k;P) 
+ \frac{r_A P_\mu}{P^2 + m_\pi^2} \Gamma_\pi^j(k;P)\,.
\end{eqnarray}

In the chiral limit, $\hat m = 0$, the AV-WTI
\begin{eqnarray}
\label{avwtich}
-i P_\mu \Gamma_{5\mu}^j(k;P)\, \frac{Z_2}{Z_A} & = & 
S^{-1}(k_+)\gamma_5\frac{\tau^j}{2}
+  \gamma_5\frac{\tau^j}{2} S^{-1}(k_-) \,,
\end{eqnarray}
where the ratio $Z_2/Z_A$ is a finite, renormalisation-group-invariant
quantity~\cite{PW68}, and hence we are free to choose
\begin{equation}
Z_A(\mu,\Lambda) = Z_2(\mu,\Lambda)\,.
\end{equation}
If one assumes $m_\pi^2=0$ in (\ref{truavv}), substitutes it into the
left-hand-side (l.h.s.) of (\ref{avwtich}) along with (\ref{sinvp}) on the
right, and equates terms of order $(P_\nu)^0$ and $P_\nu$, one obtains the
chiral-limit relations
\begin{eqnarray}
\label{bwti} 
r_A E_\pi(k;0)  &= &  B(k^2)\,, \\
 F_R(k;0) +  2 \, r_A F_\pi(k;0)                 & = & A(k^2)\,, \\
G_R(k;0) +  2 \,r_A G_\pi(k;0)    & = & 2 A^\prime(k^2)\,,\\
\label{gwti} 
H_R(k;0) +  2 \,r_A H_\pi(k;0)    & = & 0\,.
\end{eqnarray}
In perturbation theory, $B(k^2) \equiv 0$ in the chiral limit.  The
appearance of a $B(k^2)$-nonzero solution of (\ref{gendse}) in the chiral
limit signals D$\chi$SB: one has {\it dynamically generated} a quark mass
term in the absence of a seed-mass.  Equations (\ref{bwti})-(\ref{gwti}) show
that when chiral symmetry is dynamically broken: 1) the homogeneous,
isovector, pseudoscalar BSE has a massless, $P^2=0$, solution; 2) the BSA for
the massless bound state has a term proportional to $\gamma_5$ alone, with
$E_\pi(k;0)$ completely determined by the scalar part of the quark self
energy, in addition to other pseudoscalar Dirac structures, $F_\pi$, $G_\pi$
and $H_\pi$, that are nonzero in general; 3) the axial-vector vertex is
dominated by the pion pole for $P^2\simeq 0$.  The converse is also true.
Dynamical chiral symmetry breaking is therefore a sufficient and necessary
condition for the appearance of a massless pseudoscalar bound state of
very-massive constituents that dominates the axial-vector vertex.

For $\hat m \neq 0$, the AV-WTI identity is
\begin{equation}
\label{avwti}
-i P_\mu \Gamma_{5\mu}^j(k;P)  = S^{-1}(k_+)\gamma_5\frac{\tau^j}{2}
+  \gamma_5\frac{\tau^j}{2} S^{-1}(k_-) 
- 2\,m_{(\mu)} \,\Gamma_5^j(k;P) \frac{Z_4}{Z_P},
\end{equation}
where the isovector, pseudoscalar vertex is given by
\begin{eqnarray}
\label{genpve}
\left[\Gamma_{5}^j(k;P)\right]_{tu} & = & 
Z_P\,\left[\gamma_5 \frac{\tau^j}{2}\right]_{tu} \,+ 
\int^\Lambda_q \,
\left[ \chi_5^j(q;P)\right]_{sr}
K^{rs}_{tu}(q,k;P)\,,
\end{eqnarray}
with $\chi_5^j(q;P) \equiv S(q_+) \Gamma_{5}^j(q;P) S(q_-)$.  In
(\ref{genpve}), the ratio $Z_4/Z_P$ is a renormalisation-group-invariant
quantity and hence we are free to choose
\begin{equation}
Z_P(\mu,\Lambda) = Z_4(\mu,\Lambda)\,.
\end{equation}

As argued in connection with (\ref{genave}), the solution of (\ref{genpve})
has the general form
\begin{eqnarray}
\label{genpvv}
i \Gamma_{5 }^j(k;P) & = &
\frac{\tau^j}{2} \gamma_5 
\left[ i E_R^P(k;P) + \gamma\cdot P \, F_R^P  
+ \gamma\cdot k \,k\cdot P\, G_R^P(k;P) 
\right. \\
& & \nonumber
\left.+ 
\sigma_{\mu\nu}\,k_\mu P_\nu \,H_R^P(k;P)
\right]
+  \frac{ r_P }{P^2 + m_\pi^2} \Gamma_\pi^j(k;P)\,,
\end{eqnarray}
where $E_R^P$, $F_R^P$, $G_R^P$ and $H_R^P$ are regular as $P^2\to -m_\pi^2$;
i.e., the isovector, pseudoscalar vertex also receives a contribution from
the pion pole.  In this case equating pole terms in the AV-WTI entails
\begin{equation}
\label{gmora}
r_A \,m_\pi^2 = 2\,m_{(\mu)} \,r_P\,.
\end{equation}
The question arises: ``What are the residues $r_A$ and $r_P$?''

To address this we note that (\ref{genave}) can be rewritten as
\begin{eqnarray}
\label{genaveM}
\lefteqn{ \frac{1}{Z_2} \, \left[\Gamma_{5\mu}^j(k;P)\right]_{tu}  = 
}\\
&& \nonumber  
\left[\gamma_5\gamma_\mu \frac{\tau^j}{2}\right]_{tu} \,+ 
\int^\Lambda_q \,
\left[ S(q_+) \frac{\tau^j}{2} \gamma_5 \gamma_\mu S(q_-)\right]_{sr}
M^{rs}_{tu}(q,k;P)\,,
\end{eqnarray}
where $M$ is the renormalised, fully-amputated quark-antiquark scattering
amplitude: $M = K + K (SS) K + \ldots$~\cite{dserev}, which can be decomposed
as:
\begin{eqnarray}
\label{Mexpand}
M^{rs}_{tu}(q,k;P)  
& = & 
 \left[\bar\Gamma_\pi^\ell(q;-P)\right]_{rs}
        \, \frac{1}{P^2+m_\pi^2} \, \left[\Gamma_\pi^\ell(k;P)\right]_{tu} 
         +  R^{rs}_{tu}(q,k;P)\,,
\end{eqnarray}
where $\bar \Gamma_\pi(k,-P)^{\rm T} = C^{-1} \Gamma_\pi(-k,-P) C$, with
$C=\gamma_2 \gamma_4$, the charge conjugation matrix, and
$R^{rs}_{tu}(q,k;P)$ is regular as $P^2\to -m_\pi^2$.  The fact that here the
pion pole has unit residue follows from the canonical normalisation of the
BSA~\cite{LS69}:
\begin{eqnarray}
\label{pinorm}
\lefteqn{ 2 \delta^{ij} P_\mu = 
\int^\Lambda_q \left\{\rule{0mm}{5mm}
{\rm tr} \left[ 
\bar\Gamma_\pi^i(q;-P) \frac{\partial S(q_+)}{\!\!\!\!\!\!\partial P_\mu} 
\Gamma_\pi^j(q;P) S(q_-) \right]
\right. + 
} \\
& & \nonumber \left.  
\;\;\;\;\;\;\;\;\;\;\;\;\;\;\;\;\;\;\;
 {\rm tr} \left[ 
\bar\Gamma_\pi^i(q;-P) S(q_+) \Gamma_\pi^j(q;P) 
        \frac{\partial S(q_-)}{\!\!\!\!\!\!\partial P_\mu}\right]
\rule{0mm}{5mm}\right\} +  \\
& & \nonumber
\;\;\;\;\;\;\;\;\;\;
 \int^\Lambda_{q}\int^\Lambda_{k} \,[\bar\chi_\pi^i(q;-P)]_{sr} 
\frac{\partial K^{rs}_{tu}(q,k;P)}
{\!\!\!\!\!\!\!\!\!\!\!\!\partial P_\mu}\, 
[\chi_\pi^j(k;P)]_{ut}\,.
\end{eqnarray}

Substituting (\ref{truavv}) on the l.h.s. of (\ref{genaveM}),
(\ref{Mexpand}) on the right, and equating residues at the pion pole, one
obtains
\begin{eqnarray}
\label{caint}
\delta^{ij} r_A P_\mu = 
Z_2\int^\Lambda_q\,
{\rm tr}\left[\frac{\tau^i}{2} \gamma_5 \gamma_\mu 
S(q_+) \Gamma_\pi^j(q;P) S(q_-)\right]\,.
\end{eqnarray}
The factor of $Z_2$ on the right-hand-side (r.h.s.) is necessary to ensure
that $r_A$ is independent of the renormalisation point, regularisation
mass-scale and gauge parameter: recall that $Z_2\equiv 1$ at one-loop in
Landau gauge.

The renormalised axial-vector vacuum polarisation is 
\begin{eqnarray}
\label{avvp}
\Pi_{{\rm w}\mu\nu}^{ij}(P) 
& = &  
\delta^{ij}(Z_3^{\rm w} - 1) (\delta_{\mu\nu} P^2 - P_\mu P_\nu) 
-Z_2 g_{\rm w}^2 
\int^\Lambda_q \,
{\rm tr}\left[ \sfrac{\tau^i}{2}\gamma_5 \gamma_\mu \chi_{5\nu}^j(q;P)\right]\,
\end{eqnarray}
where $Z_3^{\rm w}$ is the weak-boson wave-function renormalisation constant
and $g_{\rm w}$ is the electroweak coupling.  The pion leptonic decay
constant, $f_\pi$, is obtained from the pion-pole contribution to this vacuum
polarisation:
\begin{eqnarray}
\label{piren}
\Pi_{{\rm w}\mu\nu}^{ij}(P) & = &
\delta^{ij}\,\left(\delta_{\mu\nu} P^2 - P_\mu P_\nu\right)\,
\left[ \Pi(P^2) + g_{\rm w}^2 \, f_\pi^2 \frac{1}{P^2+m_\pi^2}\right]\,,
\end{eqnarray}
where $\Pi(P^2)$ is regular as $P^2\to -m_\pi^2$.  Substituting (\ref{piren})
in the l.h.s. of (\ref{avvp}), (\ref{truavv}) on the right,
projecting with $(\delta_{\mu\nu} P^2 - 4 P_\mu P_\nu)$, and equating pole
residues one obtains
\begin{equation}
\label{cafpi}
r_A = f_\pi\,;
\end{equation}
i.e., {\it the residue of the pion pole in the axial-vector vertex is the pion
decay constant}$\,$.

The relationship, in the chiral limit, between the normalisation of the pion
BSA and $f_\pi$ has often been discussed.\footnote{For example,
Refs.~\protect\cite{FR96,MMY88}.  As shown in Ref.~\cite{cdrpion}, and
contrary to the suggestion in Ref.~\protect\cite{MMY88}, the integral
appearing in the calculation of $\pi^0 \to \gamma\gamma$ {\it does not}
provide an additional ``definition'' of $f_\pi$; it must yield $1/2$,
independent of the details of the model, in order to be consistent with the
anomalous Ward-Takahashi identity for the isosinglet axial-vector vertex.}
Consider that if one chooses to normalise $\Gamma_\pi^j$ such that
$E_\pi(0;0)= B(0)$, and defines the BSA so normalised as $\Gamma_{\pi
{N_\pi}}^j(k;P)$, then the r.h.s. of (\ref{pinorm}), evaluated with
$\Gamma_\pi^j \to \Gamma_{\pi {N_\pi}}^j$, is equal to $2 P_\mu N_\pi^2$,
where $N_\pi$ is a dimensioned constant.  Using (\ref{bwti})-(\ref{gwti}) it
is clear that in the chiral limit
\begin{equation}
\label{npifpi}
N_\pi = f_\pi\,.
\end{equation}  
However, in model studies to date, this result is not obtained {\it unless}
one assumes $A(k^2)\equiv 1$.  It follows that any kernel which leads, via
(\ref{gendse}), to $A(k^2)\equiv 1$ must also yield, $F_\pi \equiv 0 \equiv
G_\pi\equiv H_\pi$, if it preserves the AV-WTI.  In realistic model studies,
where $A(k^2) \not \equiv 1$, the difference between the values of $N_\pi$
and $f_\pi$ is an artefact of neglecting $F_\pi$, $G_\pi$ and $H_\pi$ in
(\ref{genpibsa})~\cite{dserev}.

To determine $r_P$, one rewrites (\ref{genpve}) as
\begin{eqnarray}
\label{genpveM}
\frac{1}{Z_4} \left[\Gamma_{5}^j(k;P)\right]_{tu} & = &
\left[\gamma_5\frac{\tau^j}{2}\right]_{tu} \,+ \int^\Lambda_q 
\left[ S(q_+) \frac{\tau^j}{2} \gamma_5 S(q_-)\right]_{sr}
M^{rs}_{tu}(q,k;P)\,.
\end{eqnarray}
Substituting (\ref{genpvv}) in the l.h.s. of (\ref{genpveM}),
(\ref{Mexpand}) on the right, and equating residues at the pion pole, one
obtains 
\begin{equation}
\label{cpres}
i \delta^{ij} r_P = Z_4\,\int^\Lambda_q {\rm tr}\left[
\sfrac{1}{2} \tau^i \gamma_5 S(q_+) \Gamma_\pi^j(q;P) S(q_-)\right]\,.
\end{equation}
The factor $Z_4$ on the r.h.s. depends on the gauge parameter, the
regularisation mass-scale and the renormalisation point.  This dependence is
exactly that required to ensure that: 1) $r_P$ is finite in the limit
$\Lambda\to \infty$; 2) $r_P$ is gauge-parameter independent; and 3) the
renormalisation point dependence of $r_P$ is just such as to ensure that the
r.h.s. of (\ref{gmora}) is renormalisation point {\it independent}.  This is
obvious at one-loop order, especially in Landau-gauge where $Z_2\equiv 1$ and
hence $Z_4 = Z_m$.

In the chiral limit, using (\ref{genpibsa}), (\ref{bwti})-(\ref{gwti}) and
(\ref{cafpi}), (\ref{cpres}) yields
\begin{equation}
\label{cbqbq}
\begin{array}{lcr}
\displaystyle
r_P^0  =  \frac{1}{f_\pi}\, \langle \bar q q \rangle_\mu^0 \,
, & & 
\displaystyle
\langle \bar q q \rangle_\mu^0 \equiv  
Z_4(\mu,\Lambda)\, N_c \int^\Lambda_q\,{\rm tr}_D
        \left[ S_{\hat m =0}(q) \right]\,;
\end{array}
\end{equation}
$ \langle \bar q q \rangle_\mu^0 $ is the chiral-limit {\it vacuum quark
condensate}.  It is renormalisation-point dependent but independent of the
gauge parameter and the regularisation mass-scale.  (\ref{cbqbq})
demonstrates that {\it the chiral-limit residue of the pion pole in the
pseudoscalar vertex is}$\,$ $ \langle \bar q q \rangle_\mu^0 /f_\pi$.

Using (\ref{cafpi}) and (\ref{cbqbq}), (\ref{gmora}) yields
\begin{equation}
\label{gmor}
f_\pi^2 m_\pi^2 = 2 \, m_{(\mu)}\, \langle \bar q q \rangle_\mu^0 + {\rm O}(\hat
m^2)\,. 
\end{equation}

In general, (\ref{gmora}) is the statement
\begin{equation}
\label{gmorgen}
f_\pi^2 \, m_\pi^2 =  2 \, m_{(\mu)} \, \langle \bar q q \rangle_\mu^{\hat m}\,,
\end{equation}
where we have introduced the {\it notation} $\langle \bar q q
\rangle_\mu^{\hat m} \equiv f_\pi r_P$.\footnote{We emphasise that, for $\hat
m \neq 0$, the r.h.s. of (\ref{gmora}) is {\it not} a difference of vacuum
quark condensates; a phenomenological assumption often employed.}  This
result is qualitatively equivalent to that obtained in
Ref.~\cite{miransky93}.  It is exact, with what is commonly known as the
Gell-Mann--Oakes--Renner relation, (\ref{gmor}), an obvious corollary.  The
extension of (\ref{gmorgen}) to $SU(N_f\geq 3)$, and $m_u \neq m_d$, is
relatively straightforward: (\ref{caint}) and (\ref{cpres}) remain correct,
apart from obvious modifications associated with the flavour-dependence of
the Schwinger functions, and possible flavour-dependence of the
renormalisation scheme; and the analogues of (\ref{gmor}) provide a good
approximation for $SU(N_f= 3)$, when $\hat m_f \lsim \Lambda_{\rm QCD}$.

As remarked above, the quark-antiquark scattering kernel appears in each of
(\ref{genbse}), (\ref{gendse}), (\ref{genave}) and (\ref{genpve}), and any
practical, quantitative calculation of pion (and other hadron) observables
will involve a direct or implicit truncation of $K$.  Our discussion
indicates the crucial nature of chiral symmetry and its dynamical breakdown
in connection with the pion.  This entails that in developing a tractable
truncation for quantitative calculations of pion observables, it is necessary
to ensure that (\ref{avwti}) is preserved.  The simplest truncation to do
this is the ``rainbow-ladder approximation'', in which
\begin{equation}
K^{rs}_{tu}(q,k;P) = - g^2\,D_{\mu\nu}(k-q)\,
        \left(\gamma_\mu\frac{\lambda^a}{2}\right)_{tr}\,
        \left(\gamma_\nu\frac{\lambda^a}{2}\right)_{su}
\end{equation}
in (\ref{genbse}), (\ref{genave}) and (\ref{genpve}), and $ Z_1
\Gamma_\nu(q,p) = \gamma_\nu $ in (\ref{gendse}), with a model form for
$D_{\mu\nu}(k)$ chosen consistent with Refs.~\cite{gpropir}.  This truncation
has been used extensively~\cite{dserev} and provides a quantitatively
reliable description of many pseudoscalar, vector and axial-vector meson
observables {\it without} fine-tuning.  

We note that in ``rainbow-ladder approximation'' $\partial_\mu^P K(q,k;P)
\equiv 0$ in (\ref{pinorm}) and it is a simple exercise to demonstrate
explicitly the chiral limit equality: $N_\pi = f_\pi$.  One
method~\cite{JJ73} is to use (\ref{genave}) to eliminate $Z_2 (\tau^i/2)
\gamma_5 \gamma_\mu$ on the r.h.s. of (\ref{caint}) in favour of the
regular-part of the axial-vector vertex in (\ref{truavv}).  Integration by
parts in the expression for $N_\pi^2$, neglecting surface terms that vanish
in a translationally invariant regularisation scheme, then yields the
expected result.  In this truncation, using a straightforward modification of
the method employed in Ref.~\cite{DS79}, it is also a simple exercise to
confirm (\ref{bwti})-(\ref{gwti}) from a direct comparison of (\ref{gendse}),
in the chiral limit, with (\ref{genave}) and (\ref{truavv}).  To elucidate
the results derived herein, a detailed numerical study of a QCD-based model
using this truncation is underway.  As an example of the results to expect,
neglecting $F_\pi$, $G_\pi$ and $H_\pi$ in solving the pion BSE leads to a
reduction in the eigenvalue of $\approx 40$\%; i.e., an underestimation of
$m_\pi$ by $20$\%.

A systematic extension of ``rainbow-ladder approximation'' is introduced in
Ref.~\cite{brs96}.  Therein it is shown that higher order contributions to
the kernel cancel approximately, order-by-order, so that ``rainbow-ladder
approximation'', as defined herein, provides a good estimate of the
properties of the pion (and other flavour-octet pseudoscalar and vector
mesons); i.e., for these mesons, there are no large corrections to the
rainbow-ladder results from contributions to the kernel that are neglected in
this truncation.

Herein the primary results are: (\ref{bwti})-(\ref{gwti}) and their
consequences, (\ref{cafpi}), (\ref{npifpi}), (\ref{cbqbq}) and
(\ref{gmorgen}), which entails (\ref{gmor}), and their derivation in a
model-independent manner using the DSEs.  In any model study, if one employs
a truncation that violates (\ref{avwti}), then, in general, these results
will not be recovered and one may arrive at erroneous conclusions.

To conclude, our analysis indicates that understanding the ``dichotomy'' of
the pion as both a Goldstone mode and a bound state is straightforward.  It
also unambiguously identifies the root cause of the defect in model studies
that require fine-tuning to describe the pion as a nearly-massless bound
state of very-massive constituents.  Models that do not preserve the
axial-vector Ward-Takahashi identity, (\ref{avwti}), although perhaps valid
in the regime where explicit chiral symmetry breaking effects dominate,
should not be expected to provide generally robust insight into the
qualitative features of pion observables.

We acknowledge useful conversations with A. Bender, M. C. Birse, F. T. Hawes
and J. A. McGovern.  CDR is grateful to the Department of Physics and
Mathematical Physics at the University of Adelaide for their hospitality and
support during a term as a Distinguished Visiting Scholar in which some of
this work was conducted.  This work was supported in part by the National
Science Foundation under grant no. PHY94-14291, by the Department of Energy,
Nuclear Physics Division, under contract no. W-31-109-ENG-38; and benefited
from the resources of the National Energy Research Scientific Computing
Center.


\end{document}